%% file: main.tex
\documentclass[twocolumn]{autart}    

\usepackage{graphicx}          

\usepackage{amsmath,amssymb,amsfonts}
\DeclareMathAlphabet{\mathpzc}{OT1}{pzc}{m}{it}
\usepackage{algorithm}

\usepackage{algorithmic}

\usepackage{textcomp}

\usepackage{tabularx,booktabs,ragged2e}
\newcolumntype{Y}{>{\RaggedRight\arraybackslash}X}

\usepackage{float}

\usepackage{savesym}
\savesymbol{AND}

\usepackage{amsfonts}
\newcommand{\overbar}[1]{\mkern 1.7mu\overline{\mkern-1.7mu#1\mkern-1.7mu}\mkern 1.7mu}

\newcommand\setItemnumber[1]{\setcounter{enumi}{\numexpr#1-1\relax}}

\usepackage{color,soul} 
\usepackage{comment} 
\usepackage{MnSymbol}

\usepackage{xcolor}

\usepackage{latexsym} 

\begin{document}

\setlength{\abovedisplayskip}{2pt}
\setlength{\belowdisplayskip}{2pt}

\begin{frontmatter}

\title{DMPC: A Data-and Model-Driven Approach to Predictive Control\thanksref{footnoteinfo}} 

\thanks[footnoteinfo]{This paper was not presented at any IFAC 
meeting. Corresponding author Hassan Jafarzadeh. Tel. +1-434-3289527.}

\author[UVA]{Hassan Jafarzadeh}\ead{hj2bh@virginia.edu},    
\author[ISU]{Cody Fleming}\ead{flemingc@iastate.edu}               

\address[UVA]{Department of Systems Engineering, University of Virginia, 151 Engineer's Way Charlottesville, Virginia 22904, USA}  
\address[ISU]{Department of Mechanical Engineering, Iowa State University, Ames, Iowa 50011, USA}  


\begin{abstract}                          
This work presents DMPC (Data-and Model-Driven Predictive Control) to solve control problems in which some of the constraints or parts of the objective function are known, while others are entirely unknown to the controller. It is assumed that there is an exogenous ``black box'' system, e.g. a machine learning technique, that predicts the value of the unknown functions for a given trajectory. DMPC (1) provides an approach to merge both the model-based and black-box systems; (2) can cope with very little data and is sample efficient, building its solutions based on recently generated trajectories; and (3) improves its cost in each iteration until converging to an optimal trajectory, typically needing only a few trials even for nonlinear dynamics and objectives. Theoretical analysis of the algorithm is presented, proving that the quality of the trajectory does not worsen with each new iteration, as well as providing bounds on the complexity
. We apply the DMPC algorithm to the motion planning of an autonomous vehicle with nonlinear dynamics. 

\end{abstract}

\end{frontmatter}

\section{Introduction}\label{sec:intro}
\input{introduction.tex}

\section{Problem Statement}\label{sec:preliminaries}
\input{preliminaries.tex}

\section{DMPC Approach} \label{sec:DMPC_approach}
\input{DMPC_approach}

\section{Implementation Steps} \label{sec:implementation}
\input{implementation}

\section{Example} \label{sec:example}
\input{example.tex}

\section{Conclusions}\label{sec:conclusions}
\input{conclusion.tex}

\begin{ack}                               
This work was partially supported by the NSF under grants CPS-1739333.  
\end{ack}

\bibliographystyle{plain}        
\bibliography{main.bib}           





\end{document}

%% file: introduction.tex
Traditional techniques for analyzing and developing control laws in safety-critical applications usually require a precise mathematical model of the system \cite{ames2016control,taylor2020learning}. However, there are many control applications where such precise, analytical models cannot be derived or are not readily available. System identification is a parametric model approach to such problems, mostly focusing on asymptotic error characterization or consistency guarantees, often assuming that the structure of the underlying system is known or that states are directly measurable~{\cite{matni2019self,sarkar2019finite}}. On the other hand, data-driven approaches from machine learning are used in order to address these cases in a non-parametric way and often can be successful even with no assumptions about the structure of the underlying system. Such approaches can be used to identify unmodeled dynamics in a scalable way, and with high accuracy. However, an objective that is increasingly prevalent in the literature involves merging or complementing the analytical approaches from control theory with techniques from machine learning.


Recently, techniques based on model-predictive control (MPC) have addressed this problem by first using a statistical method to estimate a mathematical model that is compatible with the data, and then using this estimated model within a nominal MPC framework to find optimal trajectories and control actions. In addition to the aforementioned system identification techniques, a popular choice is to build statistical models using Gaussian Processes (GPs) \cite{deisenroth2013gaussian,kamthe2018data}, while Regression Trees and other machine learning techniques have been used in other cases \cite{behl2016dr}. The use of GPs in the context of model-predictive control often creates highly nonlinear models, resulting in non-convex problems that are difficult to solve efficiently or online. Alternatively, approaches based on Reinforcement Learning have been applied in this setting. Model-based techniques again require a statistical method, for example, GPs or deep neural networks, to estimate transition probability distributions \cite{moerland2020model}. Model-free methods represent, informally, a trial-and-error method for identifying control policies \cite{deisenroth2011pilco,kamthe2017data}. An open question in reinforcement learning (and indeed much of the literature that uses both control theory and machine learning) involves how to guarantee that the learned policy will not violate safety or other constraints \cite{bacci2020probabilistic,isele2018safe}. In addition, sample complexity represents a challenge for all the aforementioned techniques and is a general problem in machine learning. 

This paper seeks to leverage the notion that in many applications, some aspects of the system (and environment) may be known mathematically while other aspects are unknown or represented by a so-called ``black box''. Our method attempts to utilize the capabilities of model-based (MPC) and data-driven (machine learning algorithm) approaches, and bring them together in a single framework in planning and control problems.

The paper addresses both sample complexity and online computational efficiency by dividing the state space, such that the dimensionality of the mathematical models and the data needed for statistical estimation and prediction are both reduced, while also accounting for the interconnection between these two classes of variables. Furthermore, we develop an algorithm that leverages this decoupling of variables, and efficiently focuses on a specific part of the state space that likely contains the optimal, feasible trajectory without sampling from the rest of the state space. Specifically, we assume that the dynamics of the system are available in the form of a known mathematical model, but there is an unknown function of the states and control inputs of the system that affects the performance index or feasible solution space. It is also assumed that the unknown aspects of the system or environment can be predicted/measured for a given system trajectory, e.g. by a ``black box''.

Our technique is based on notions from Iterative Learning Control (ILC). ILC is attractive because it can ``learn’’ through repeated trials to converge to better solutions~\cite{wang2009survey}. The concept of ILC has recently been extended to a framework that does not require a reference signal~\cite{rosolia2018learning,rosolia2017robust}, although this approach still assumes that initial conditions, constraints, and costs remain consistent at each iteration. Although the aforementioned techniques have several nice qualities, e.g. no need for a reference signal or known cost function, they (a) assume a repetitive setting and (b) generally do not apply to so-called ``black box’’ variables. We borrow from ILC concepts but generalize to non-repetitive or non-iterative tasks, where a controller needs to make real-time decisions in novel environments. Furthermore, our approach works when the dynamics are unknown for at least some aspects of the system or environment. The approach leverages machine learning and MPC to predict the behavior of the black-box and mathematically modeled components of the system, respectively, incorporating both into a technique called Data- and Model-driven Predictive Control (DMPC). DMPC works without a reference signal and – for a subset of the state or cost variables – completely unknown dynamics; furthermore, DMPC can work with an unknown cost function. We prove that DMPC is recursively feasible at each iteration of the algorithm, and the generated trajectories will not worsen at each iteration. This algorithm needs only a few iterations to converge to a locally optimal solution and is computationally efficient, even for nonlinear system dynamics. We also demonstrate the performance of the algorithm with an application to a motion planning problem with nonlinear dynamics in a totally unknown environment. 

%% file: preliminaries.tex
In this section, a formal definition of the problem is presented. Consider the dynamical system:
\begin{equation}
    x_{t+1} = f\left(x_t,u_t \right), \label{eq:nldynamics}
\end{equation}
where $x\in\mathbb{R}^n$ and $u\in\mathbb{R}^m$ are the system states and control inputs, respectively, and $f:\mathbb{R}^n \times \mathbb{R}^m \rightarrow \mathbb{R}^n$ is a known and in general nonlinear map which assigns the successor state $x_{t+1}$ to state $x_t$ and control input $u_t$. In this paper we address the following infinite time optimal control problem to find an optimal trajectory from an initial state $x_S$ to final state $x_F$ within the feasible state vector space $\mathcal{X}$ and control vector space $\mathcal{U}$:
\begin{subequations}\label{eq:main_model_generic_entire_model}
    \begin{align}
        \mathcal{J}_{0\to\infty}(x_S) &= \min_{u_0,u_1,\dots}\sum_{t=0}^{\infty}\left[ h\left(x_t , u_t\right) + \hat{z}\left(x_t, u_t\right)\right]\label{eq:main_model_generic-objective}\\
        \text{s.t.}\ & x_{t+1} = f\left(x_t,u_t \right) \quad \forall t \ge 0 \label{eq:main_model_dynamic1}\\
        & x_0 = x_S \label{eq:main_model_init} \\
        & x_t \in \mathcal{X}, \quad u_t \in \mathcal{U}\quad \forall t \ge 0, \label{eq:main_model_xu_consts0}
    \end{align}
\end{subequations}
where (\ref{eq:main_model_dynamic1}) and (\ref{eq:main_model_init}) are the system dynamics and the initial conditions, and (\ref{eq:main_model_xu_consts0}) are the state and input constraints. The cost function involves two different stage costs. \textit{i)} $h()$: a \textbf{known function} that can be defined by a precise mathematical model, often based on first principles from physics. We call this a ``\textit{model-driven}" function. The traditional cost function of MPC, containing quadratic terms to drive the state of the system to an equilibrium point and to penalize the applied control input, consists of model-driven functions. 
\textit{ii)} $\hat{z}()$: an \textbf{unknown function} to the controller. A mathematical model cannot be defined for this type of stage cost (or at least it is too expensive to derive such a function and solve the resulting optimization model), but it affects the overall cost function. It is assumed that, given the inputs, the controller has access to the output of this function. 
Improving an aircraft's flight safety under the presence of turbulence can be modeled as (\ref{eq:main_model_generic_entire_model}), where the behavior, location, and prediction of turbulent air comes from an unknown function (unknown to the controller). Another example involves connected autonomous vehicles (CAVs) ~\cite{9029830,soltanaghaei2019characterizing}, in which the unknown function is a model of the wireless channel and can be predicted by e.g. recurrent neural networks~\cite{manh2018scene,liu2019deepvm}.

It is assumed that the model-driven stage cost $h(\cdot,\cdot)$ in equation (\ref{eq:main_model_generic-objective}) is continuous and satisfies $h\left(x_F,0\right)=0,$ \[ h\left(x_t , u_t\right)\succ 0 \ \forall x_t \in \mathbb{R}^n \backslash \left\{x_F\right\}, u_t \in \mathbb{R}^m \backslash \left\{0\right\}\] where the final state $x_F$ is a feasible equilibrium for the unforced system (\ref{eq:nldynamics}), $f(x_F,0)=x_F$. In the second term of the cost function, $\hat{z}()$ is considered to be positive definite and unknown for the controller, $\hat{z}:\mathbb{R}^{n}\times\mathbb{R}^{m}\to\mathbb{R}^{+}$. 
There is an exogenous data-driven system acting as a black box, such as Long short-term memory (LSTM) that calculates $\hat{z}$, given $x_t$ and $u_t$. Also, we assume that the condition $\hat{z}\left(x_F, 0 \right) =0$ is held in the equilibrium point $x_F$.

In the case that an unknown inequality is imposed as a constraint to the model rather than a penalty in the cost function, we can use a barrier function to transform it to model (\ref{eq:main_model_generic_entire_model}). If we write these constraints as $\hat{y}\left(x_t, u_t \right) \leqslant 0, \quad \forall t \geqslant 0,$
the barrier function can be defined as 
\begin{equation} \label{eq:barrier_function}
  \hat{z}\left(x_t, u_t\right)=
  \begin{cases}
  -\frac{1}{\hat{y}\left(x_t, u_t\right)}                & \text{if} \ \hat{y}\left(x_t, u_t\right) < 0\\
  \infty                 & o.w.
  \end{cases}
\end{equation}
in the exogenous data-driven system, where the controller will receive the value of $\hat{z}()$ calculated from equation (\ref{eq:barrier_function}) and then considers this value as a prediction for the unknown cost in the performance index shown in model (\ref{eq:main_model_generic_entire_model}). Therefore, the problem involves generating an optimal sequence of control inputs that steers the system (\ref{eq:nldynamics}) from the initial state $x_S$ to the equilibrium point $x_F$ such that the cost function of optimal control problem (\ref{eq:main_model_generic_entire_model}), $\mathcal{J}_{0\to\infty}(x_S)$ -- which is a combination of a known stage cost $h()$, and unknown stage cost $\hat{z}()$ functionals -- achieves the minimum value.


At each time step of a (perhaps previously unseen) control task, the approach uses an iterative scheme, where it learns from each iteration and optimizes model (\ref{eq:main_model_generic_entire_model}) without explicitly determining the unknown function $\hat{z}()$. At iteration $j$, the following vectors collect the inputs applied to the system (\ref{eq:nldynamics}) and the corresponding state evolution from initial state $x_S$ to the equilibrium point $x_F$:
\begin{subequations}\label{eq:jth-iteration}
    \begin{align}
      \textbf{x}^{*,j} &= [ x_0^{j}, x_1^{*,j}, \dots , x_t^{*,j}, \dots, x_F]\label{eq:state1}\\
      \textbf{u}^{*,j} &= [ u_0^{*,j}, u_1^{*,j}, \dots , u_t^{*,j}, \dots] \label{eq:input1}.
    \end{align}
\end{subequations}
In (\ref{eq:jth-iteration}), the optimal values of system state and the control input obtained at time $t$ and iteration $j$ are denoted by $x_t^{*,j}$ and $u_t^{*,j}$, respectively. Also, we assume that at each $j^\text{th}$ iteration, the trajectories start from the same initial condition $x_0^j=x_{S},\quad \forall j \ge 0$.

%% file: DMPC_approach.tex
This section describes the DMPC approach to obtain vectors(\ref{eq:jth-iteration}) as a sub-optimal solution for the infinite time optimal control problem (\ref{eq:main_model_generic_entire_model}).
We begin with the following assumption, as the DMPC algorithm is designed such that, starting from a given initial trajectory, it converges to the optimal solution (trajectory) repetitively.

\textit{Assumption 1}: Similar to the iterative learning control methods~{\cite{rosolia2018learning,rosolia2017robust},} it is assumed that there exists an initial feasible trajectory $\textbf{x}^0$ for the infinite time optimal control problem (\ref{eq:main_model_generic_entire_model})  from the initial state, $x_S$, to the equilibrium point, $x_F$, at the first iteration but with no assumptions on optimality.

\begin{figure}[bt!]
    \centering
    \hspace{-10pt}\includegraphics[width=1\linewidth]{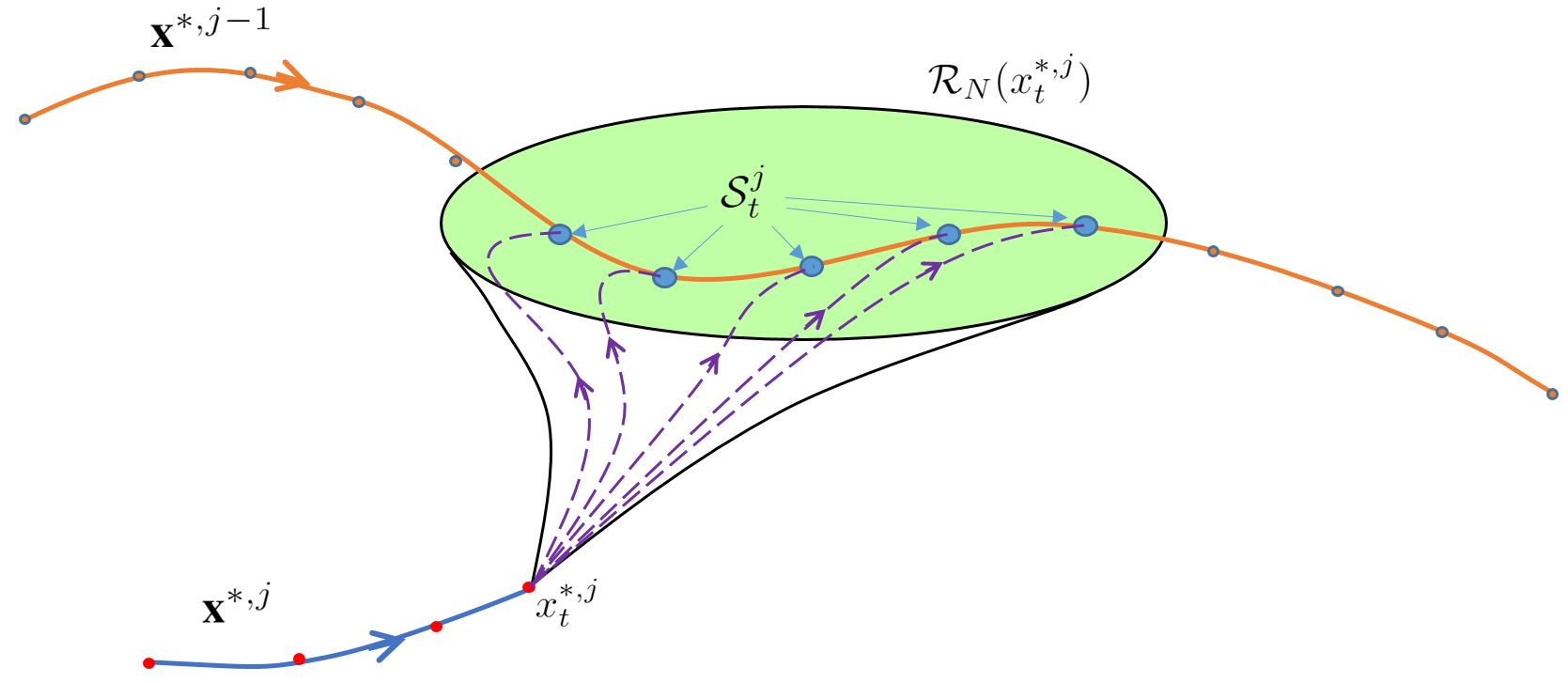}
    \caption{The green area shows the $N$-step reachable set, $\mathcal{R}_N(x_t^{*,j})$, from current state, $x_t^{*,j}$. Controllable set $\mathcal{S}_t^j$ is illustrated by large blue dots and dashed purple line segments are the optimal trajectories from current state to available states in controllable set $\mathcal{S}_t^j$.}
    \label{fig:controllable_set}
\end{figure}

In addition, the concept of cost-to-go is defined for each state in a complete trajectory as the minimum cost of reaching the equilibrium point $x_F$ from the current state. The algorithm records the last successful complete trajectory (i.e. from initial state $x_s$ to the equilibrium point $x_F$), $\textbf{x}^{*,j-1}$, and assigns to every state in this set a cost-to-go value obtained at iteration $j-1$,
\[\textbf{q}^{j-1} = [ q^{j-1}(x_S), \dots , q^{j-1}(x_t^{*,j-1}), \dots , q^{j-1}(x_F)].\] The cost of following the trajectory obtained at iteration $j-1$ from state $x_t^{*,j-1}$ to final state $x_F$ can be defined as:
\[ q^{j-1}(x_t^{*,j-1}) = \mathcal{J}_{t\to\infty}^{*,j-1}(x_t^{*,j-1}), \quad \forall t \ge 0. \]
The main approach of DMPC is generating a full trajectory from $x_S$ to $x_F$ at iteration $j$, $\textbf{x}^{*,j}$, based on the full trajectory generated at iteration $j-1$, $\textbf{x}^{*,j-1}$. The full trajectory $\textbf{x}^{*,j}$ is built iteratively from the initial state $x_S$ to the final state $x_F$. At each time step $t$ of iteration $j$, DMPC finds the optimal control input, $\textbf{u}_{t:t+N|t}^{j}$, and associated trajectory, $\textbf{x}_{t:t+N|t}^{j}$
\begin{subequations} \label{eq: dmpc_ones_step_RHC}
    \begin{align}
    \textbf{x}_{t:t+N|t}^{j} = [x_{t}^{*,j}, \dots, x_{t+N|t}^{j}] \label{eq: dmpc_ones_step_RHC_state}\\
    \textbf{u}_{t:t+N|t}^{j} = [u_{t|t}^{j}, \dots, u_{t+N-1|t}^{j}].
    \end{align}
\end{subequations}
Where $x_{t}^{*,j} = x_{t|t}^{j}$ is the current state of the system, which is considered as the optimal state of the trajectory at iteration $j$ at time $t$. DMPC selects the last state in (\ref{eq: dmpc_ones_step_RHC_state}), $x_{t+N|t}^{j}$, from a special set that results in a recursive feasibility guarantee. 
At iteration $j$, DMPC is designed by repeatedly solving a finite time optimal control problem in a receding horizon fashion to obtain state and control input vectors (\ref{eq: dmpc_ones_step_RHC}). In the state vector (\ref{eq: dmpc_ones_step_RHC_state}), the last state, $x_{t+N|t}^{j}$, is enforced to be selected from set $\mathcal{S}_t^j$, that is
\begin{equation} \label{eq:candidate_termina_states_set_0}
    \mathcal{S}_t^j = \Big(\bigcup_{t=0}^{\infty}{x}_t^{*,j-1} \Big) \cap \mathcal{R}_N(x_t^{*,j}).
\end{equation}
The first term in equation (\ref{eq:candidate_termina_states_set_0}) is the set of all the states in the most recently generated full trajectory (iteration $j-1$), $\textbf{x}^{*,j-1}$, and the second term is \textit{N-step reachable set} from state $x_{t}^{*,j}$. All the states in trajectory $\textbf{x}^{*,j-1}$ are a member of \textit{control invariant set} $\mathcal{C} \subseteq \mathcal{X}$, because, for every point in the set, there exists a feasible control action in input vector $\textbf{u}^{*,j-1}$, that satisfies the state and control constraints and steers the state of the system (\ref{eq:nldynamics}) toward the equilibrium point $x_F$. Therefore, forcing the controller to select the terminal state $x_{t+N|t}^{j}$ from the set $\mathcal{S}_t^j$ keeps the state of the system in set $\mathcal{C}$ for time steps beyond the time horizon $N$ \cite{firoozi2018safe}, i.e.
\begin{equation}
    \text{if} \quad x_{t+N}^j \in \mathcal{C} \Rightarrow x_{t+N+k}^j \in \mathcal{C} \quad \forall k>0,
\end{equation}
On the other hand, trajectory $\textbf{x}_{t:t+N|t}^{j}$ drives the system (\ref{eq:nldynamics}) from state $x_t^{*,j}$ to one of the states in set $\mathcal{S}_t^j$ in $N$ time steps (see Figure \ref{fig:controllable_set}). Therefore, $\mathcal{S}_t^j$ is a subset of the control invariant set and $N$-step reachable set, making the state $x_{t}^{*,j}$ a subset of the maximal stabilizable set. Intuitively, this guarantees the constraint satisfaction and feasibility for all time steps ($t \geqslant 0$) (the feasibility will be proven in  \textit{Theorem} (\ref{theorem:feasibility_proof})). This means that constraint satisfaction at time steps beyond the time horizon does not depend on the length of the time horizon, and $N$ can be picked freely; in this work we will select $N$ to be small to speed up the algorithm. We denote each state in set $\mathcal{S}_t^j$ by $\textit{s}_t^{i,j}, \forall i \in \{1,\dots,|\mathcal{S}_t^j| \}$.

\subsection{Algorithmic Details}

To find the (local) optimal trajectory  $\textbf{x}_{t:t+N|t}^{j}$ in (\ref{eq: dmpc_ones_step_RHC}), DMPC generates two trajectories $\overbar{\textbf{x}}_{t:t+N|t}^j$ and $\textbf{x}_{t:t+N|t-1}^{j}$, and selects the best of them based on their cost as $\textbf{x}_{t:t+N|t}^{j}$. We now explain how these two trajectories are built.

\textit{i)} The \textit{first} trajectory generated by DMPC is $\overbar{\textbf{x}}_{t:t+N|t}^j$, illustrated by a solid black trajectory in Figure~{\ref{fig:sub_trajs_DMPC}}. This trajectory is the state vector associated with the optimal control input $\overbar{\textbf{u}}_{t:t+N|t}^j$ obtained from the following optimization model over all the candidate terminal states that are reachable in $N$ time steps from the current state $x_t^{*,j}$, see equation (\ref{eq:candidate_termina_states_set_0}). This set of terminal states is depicted by big blue points in Figure \ref{fig:controllable_set} and indexed by $i\in\{1,\dots,|\mathcal{S}_t^j|\}$ in the following term
\begin{equation} \label{eq:main_DMPC_concise_model}
    \overbar{\textbf{u}}_{t:t+N|t}^j = \operatorname*{argmin}_{\textbf{u}_{t:t+N|t}^{i,j}} \Big\{\mathcal{J}_{t\to t+N}^{i,j}(x_t^{*,j}), \ \forall i\Big\}
\end{equation}
where $\mathcal{J}_{t\to t+N}^{i,j}(x_t^{*,j})$ is the predicted overall cost (i.e. summation of both the model-based $\sum_{}^{}h(.)$ and black-box $\sum_{}^{}\hat{z}(.)$ costs) due to the system following the control input $\textbf{u}_{t:t+N}^{i,j}$ to reach the terminal state $x_{t+N|t}^{i,j} = s_t^{i,j}$. To simplify the mathematical notations, we will use $\hat{z}_{k|t}^{i,j}$ to show the predicted value of the unknown function following the control input $\textbf{u}_{t:t+N}^{i,j}$, instead of $\hat{z}(x_{k|t}^{i,j}, u_{k|t}^{i,j})$. Then the value of $\mathcal{J}_{t\to t+N}^{i,j}(x_t^{*,j})$ can be defined as:
\begin{equation} \label{eq:seperated_model_data_cost}
    \mathcal{J}_{t\to t+N}^{i,j}(x_t^{*,j}) = J_{t\to t+N}^{i,j}(x_t^{*,j}) + \sum_{k=t}^{t+N-1}\hat{z}_{k|t}^{i,j}.
\end{equation}
To find the optimal control input $\overbar{\textbf{u}}_{t:t+N|t}^j$ in equation (\ref{eq:main_DMPC_concise_model}), we first use the following formulation to generate $\textbf{u}_{t:t+N|t}^{i,j}$  and $\textbf{x}_{t:t+N|t}^{i,j}$ from state $x_t^{*,j}$ toward terminal state $s_t^{i,j} \in \mathcal{S}_t^j$, $\forall i\in\{1,\dots,|\mathcal{S}_t^j|\}$, and calculate the cost associated with the model-based term, which is denoted by $J_{t\to t+N}^{i,j}(x_t^{*,j})$ in equation (\ref{eq:seperated_model_data_cost}):
\begin{subequations}\label{eq:DMPC_generic-objective_model}
    \begin{align}
        J_{t\to t+N}^{i,j}(x_t^{*,j}) &= \min_{\textbf{u}_{t:t+N}^{i,j}}\sum_{k=t}^{t+N-1} \ell(x_{k|t}^{i,j} , u_{k|t}^{i,j})\notag \\ & + (N+1)q^{j-1}(x_{t+N|t}^{i,j}) \label{eq:DMPC_generic-objective}\\
        \text{s.t.}\ & x_{k+1|t}^{i,j} = f(x_{k|t}^{i,j},u_{k|t}^{i,j})\quad \forall k \label{eq:DMPC_dynamic1}\\
        & x_{t|t}^{i,j} = x_t^{*,j} \label{eq:DMPC_init} \\
        & x_{t+N|t}^{i,j} = \textit{s}_t^{i,j} \label{eq:DMPC_terminal_state}\\
        & x_{k|t}^{i,j} \in \mathcal{X}, \ u_{k|t}^{i,j} \in \mathcal{U},\quad \forall k. \label{eq:DMPC_xu_consts0}
    \end{align}
\end{subequations}
In this model, the predictive controller generates the best trajectory to reach state $s_t^{i,j}$ (i.e. enforced by constraint (\ref{eq:DMPC_terminal_state})) and adds the cost to go $(N+1)q^{j-1}(x_{t+N|t}^{i,j})$ to compensate for the remaining cost from state $s_t^{i,j}$ to the final state $x_F$. We replace the stage cost $h(.,.)$ with a positive definite function $\ell(.,.)$ in the cost function \[\ell(x_{k|t}^{i,j}, u_{k|t}^{i,j}) =
    ||x_{k|t}^{i,j} - x_{t+N|t}^{i,j}||_{P}^2
    + ||u_{k|t}^{i,j}||_{R}^2,\]
where $P$ and $R$ are positive (semi)definite tuning matrices. The function $h()$ in the general optimal control problem(\ref{eq:main_model_generic_entire_model}) penalizes the controller according to the difference between the generated state $x_{k|t}^{i,j}$ and the final state $x_F$, but $\ell(.,.)$ considers the selected terminal state $x_{t+N|t}^{i,j}$ instead of $x_F$. To compensate for the remaining trajectory cost from $x_{t+N|t}^{i,j}$ to $x_F$, we add a cost-to-go $q^{j-1}(x_{t+N|t}^{i,j})$ for each $N+1$ states in the trajectory. Constraint (\ref{eq:DMPC_terminal_state}) 
enforces that the controller steers the system to a specific terminal state, $\textit{s}_t^{i,j}$.

The objective optimized by model (\ref{eq:DMPC_generic-objective_model}) does not involve the cost value coming from the black-box variables, $\sum_{k=t}^{t+N-1}\hat{z}_{k|t}^{i,j}$. However, given the trajectory $\textbf{x}_{t:t+N|t}^{i,j}$ generated by model (\ref{eq:DMPC_generic-objective_model}), the value of this unknown function can be predicted by the external black-box system and added to $J_{t\to t+N}^{i,j}(x_t^{*,j})$ to find $\mathcal{J}_{t\to t+N}^{i,j}(x_t^{*,j})$ based on equation (\ref{eq:seperated_model_data_cost}). Then, according to (\ref{eq:main_DMPC_concise_model}), between all of the trajectories that start from $x_t^{*,j}$ and reach the terminal states in set $\mathcal{S}_t^j$  which are counted by index $i$ (dashed purple trajectories in Fig. \ref{fig:controllable_set}), the trajectory that has the minimum cost value is selected and denoted $\overbar{\textbf{x}}_{t:t+N|t}^j$. This is the result for (\ref{eq:main_DMPC_concise_model}), where the input sequence $\overbar{\textbf{u}}_{t:t+N|t}^j$ produces the overall trajectory cost of $\overbar{\mathcal{J}}_{t\to t+N}^{j}(x_t^{*,j})$.


\textit{ii)} The \textit{second} trajectory generated by DMPC is $\textbf{x}_{t:t+N|t-1}^{j}$, that is illustrated by a dashed green trajectory in Figure~{(\ref{fig:sub_trajs_DMPC})}. In addition to $\overbar{\textbf{x}}_{t:t+N|t}^j$, another feasible available trajectory starting from $x_S$ to $x_F$ can be obtained from the solution of the previous time step $t-1$ at the current iteration $j$. This trajectory is generated by applying one more step of the control input, $\textbf{u}_{t-1:t+N-1|t-1}^{j}$, to the trajectory of the previous time step $t-1$ and shifting its state one time step toward the final state $x_F$ along the optimal trajectory of iteration $j-1$. This trajectory can be written as follows:
\begin{figure}[bt!]
    \centering
    \hspace{-10pt}\includegraphics[width=0.9\linewidth]{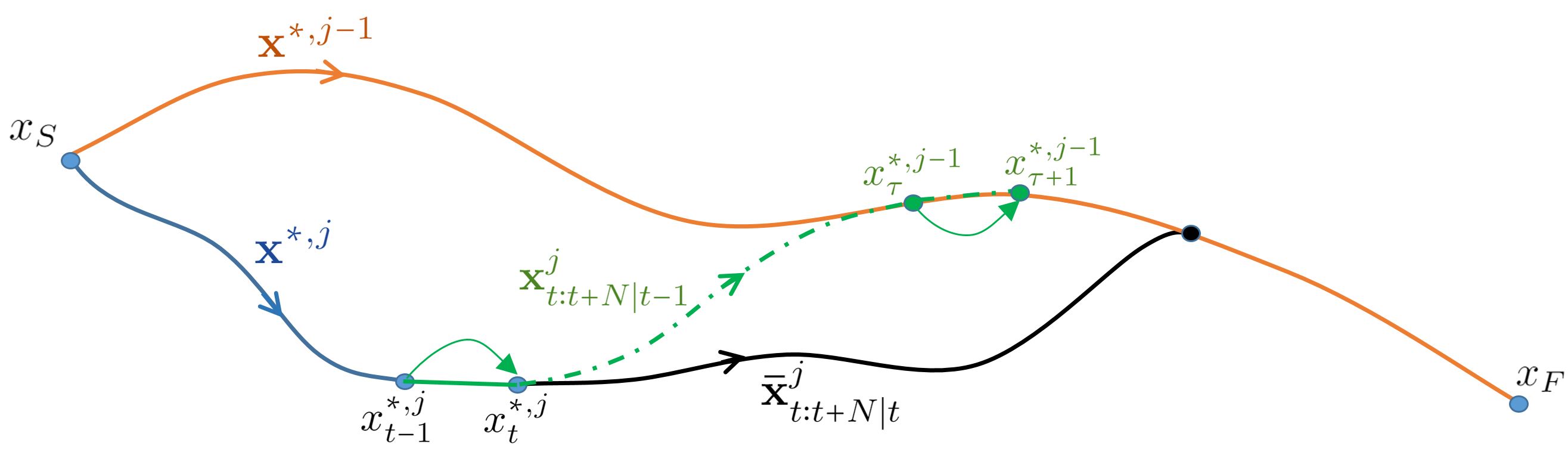}
    \caption{DMPC generates two trajectories $\overbar{\textbf{x}}_{t:t+N|t}^j$ (solid black) and $\textbf{x}_{t:t+N|t-1}^{j}$ (dashed green) at time step $t$ of iteration $j$, and selects best of them}
    \label{fig:sub_trajs_DMPC}
\end{figure}
\begin{subequations} \label{eq:trajectory_i_last_time_step}
    \begin{align}
      \textbf{x}_{t:t+N|t-1}^{j} &= \left[x_{t|t-1}^{j}, \dots,x_{t+N-2|t-1}^{j}, x_{\tau}^{*,j-1}, x_{\tau +1}^{*,j-1}\right]\\
      \textbf{u}_{t:t+N|t-1}^{j} &= \left[u_{t|t-1}^{j}, \dots,u_{t+N-2|t-1}^{j}, u_{\tau}^{*,j-1}\right].
    \end{align}
\end{subequations}
$x_{\tau}^{*,j-1}$ denotes the optimal terminal state selected from the last iteration (i.e. the last generated complete trajectory) $j-1$, and $\tau$ is the time index of this state, $x_{t+N-1|t-1}^{j} = x_{\tau}^{*,j-1}$.
The overall trajectory cost of $\textbf{x}_{t:t+N|t-1}^{j}$ is given by $\mathcal{J}_{t\to t+N|t-1}^{j}(x_{t}^{*,j})$ and is
\begin{multline} \label{eq:secondary_traj_ii}
    \mathcal{J}_{t\to t+N|t-1}^{j}(x_{t}^{*,j}) = \sum_{k=t}^{t+N-2} \left[\ell(x_{k|t-1}^{j} , u_{k|t-1}^{j}) + \hat{z}_{k|t-1}^{j}\right]\\+ Nq^{j-1}(x_{t+N-1|t-1}^{j}) + q^{j-1}(x_{\tau +1}^{*,j-1}).
\end{multline}
Finally, the best trajectory of time step $t$ and iteration $j$ ($\textbf{u}_{t:t+N|t}^{j}$ and $\textbf{x}_{t:t+N|t}^{j}$) is selected between two obtained trajectories, $\overbar{\textbf{x}}_{t:t+N|t}^j$ and $\textbf{x}_{t:t+N|t-1}^{j}$ based on their cost.
\begin{equation} \label{eq:comparision_overall_cost}
    \mathcal{J}_{t\to t+N}^{j}(x_t^{*,j}) = \min \{\overbar{\mathcal{J}}_{t\to t+N}^{j}(x_t^{*,j}), \mathcal{J}_{t\to t+N|t-1}^{j}(x_{t}^{*,j})\}.
\end{equation}
In other words, the algorithm selects between two trajectories: (a) the minimum-cost feasible trajectory from $t\to t+N$ at time step $t$ of iteration $j$, and (b) the time-shifted trajectory from $t-1\to t+N$ that leverages information from the prior time step $t-1$ of iteration $j$. After finding $\textbf{x}_{t:t+N|t}^{j}$ and $\textbf{u}_{t:t+N|t}^{j}$, the first step of its control input is applied to the system to push its state toward the equilibrium point, $u_t^{*,j} = u_{t|t}^{j}, \quad x_{t+1}^{*,j} = x_{t+1|t}^{j}$.

\subsection{Theoretical Analysis}
In the remainder of this section, we provide theoretical analyses of the algorithm for the feasibility and optimality of the generated solutions.
\begin{thm} \label{theorem:feasibility_proof}
In the DMPC scheme with given system (\ref{eq:nldynamics}), cost function (\ref{eq:main_DMPC_concise_model}), and constraints (\ref{eq:DMPC_dynamic1}) - (\ref{eq:DMPC_xu_consts0}), if there is a feasible trajectory at iteration $j-1$, DMPC is feasible at the next iteration, $j$, as well.
\end{thm}

\textit{Proof}:
To prove this theorem, first, we need to show that, given a feasible solution at time step $t-1$ of iteration $j$, DMPC is feasible for the next time step, $t$, too. The solution of DMPC at iteration $j-1$ is \[\textbf{x}^{*,j-1}=\left[x_{S}, x^{*,j-1}_{1}, \dots, x^{*,j-1}_t, \dots, x_F\right]\label{eq:traj_j-1},\] and at iteration $j$ and time step $t-1$ is: \[\textbf{x}_{t-1:t+N-1|t-1}^j = [x_{t-1}^{*,j}, x_{t|t-1}^j, \dots, x_{t+N-1|t-1}^j]\] \[\textbf{u}_{t-1:t+N-1|t-1}^j = [u_{t-1|t-1}^j,\\ u_{t|t-1}^j, \dots, u_{t+N-2|t-1}^j].\]
According to constraint (\ref{eq:DMPC_terminal_state}), DMPC selects terminal state $x_{t+N-1|t-1}^j$ from set $\mathcal{S}_{t-1}^j$ which is denoted by $s_{t-1}^{i,j}$. Because $s_{t-1}^{i,j} \in \textbf{x}^{*,j-1}$, we know that $x_{t+N-1|t-1}^j \in \textbf{x}^{*,j-1}$. Let us assume that $x_{t+N-1|t-1}^j = x_{\tau}^{*,j-1}$. Based on the assumption given in the theorem (existence of a feasible trajectory at iteration $j-1$), for every state in trajectory $\textbf{x}^{*,j-1}$ there is a feasible sequence of control actions that satisfies the constraints and steers the system toward the final state $x_F$. This feasible trajectory for state $x_{\tau}^{*,j-1}$ can be shown as: \[\textbf{x}_{\tau:\infty}^{*,j-1} = [x_{\tau}^{*,j-1}, x_{\tau+1}^{*,j-1}, \dots, x_F]\]\[\textbf{u}_{\tau:\infty}^{*,j-1} = [u_{\tau}^{*,j-1}, u_{\tau+1}^{*,j-1}, \dots].\]
Then there is at least one feasible trajectory at time step $t$ and iteration $j$ that is constructed as:
\[\textbf{x}_{t:\infty}^j = [x_{t|t-1}^j, \dots, x_{t+N-2|t-1}^j, x_{\tau}^{*,j-1}, x_{\tau+1}^{*,j-1}, \dots, x_F]\]\[\textbf{u}_{t:\infty}^j = [u_{t|t-1}^j, \dots, u_{t+N-2|t-1}^j, u_{\tau}^{*,j-1}, u_{\tau+1}^{*,j-1}, \dots].\]
This completes the proof of the statement that DMPC is feasible at time step $t$ if it is feasible at $t-1$. Also, based on \textit{Assumption} (1) and by induction we can conclude that DMPC is feasible for all iterations and time steps.   \hfill  $\filledmedsquare$

We showed that, given a feasible initial trajectory $\textbf{x}^0$, the algorithm is feasible at every time steps of different iterations. Theorem (\ref{theorem:asymptotically_stable}) proves that the algorithm will finally converge to the equilibrium point $x_F$ , and Theorem (\ref{theorem:optimality_different_iter}) proves that the performance index is non-increasing at every DMPC iteration. The next two theorems follow a similar approach to \cite{rosolia2018learning}.

\begin{thm}\label{theorem:asymptotically_stable}
In the DMPC scheme with given system (\ref{eq:nldynamics}), cost function (\ref{eq:main_DMPC_concise_model}), constraints (\ref{eq:DMPC_dynamic1}) - (\ref{eq:DMPC_xu_consts0}), and an initial feasible trajectory  
\normalfont $\textbf{x}^0$
, the equilibrium point $x_F$ is asymptotically stable at
every iteration $j \geqslant 1$.
\end{thm}

\textit{Proof}:
Let us start with writing the overall optimal trajectory cost of state $x_{t-1}^{*,j}$
\[\mathcal{J}_{t-1\to t+N-1}^{j}(x_{t-1}^{*,j}) =(N+1)q^{j-1}(x_{t+N-1|t-1}^{j})\]\[+ \sum_{k=t-1}^{t+N-2} \left[\ell(x_{k|t-1}^{j} , u_{k|t-1}^{j}) + \hat{z}_{k|t-1}^{j}\right]\]\[ = \ell(x_{t-1|t-1}^{j} , u_{t-1|t-1}^{j}) + \hat{z}_{t-1|t-1}^{j} + q^{j-1}(x_{t+N-1|t-1}^{j})+ \]\[\sum_{k=t}^{t+N-2} \ell(x_{k|t-1}^{j} , u_{k|t-1}^{j}) + Nq^{j-1}(x_{t+N-1|t-1}^{j})+q^{j-1}(x_{\tau +1}^{*,j-1})\]
where $q^{j-1}(x_{\tau+1}^{*,j-1}) = \sum_{k=\tau+1}^{\infty} \left[ h\left(x_k^{*,j-1} , u_k^{*,j-1}\right) + \hat{z}_{k}^{*,j-1}\right].$ 
Using equation (\ref{eq:secondary_traj_ii}),
\[\mathcal{J}_{t-1\to t+N-1}^{j}(x_{t-1}^{*,j}) =\mathcal{J}_{t\to t+N|t-1}^{j}(x_{t}^{*,j})\]
\[+ \ell(x_{t-1|t-1}^{j} , u_{t-1|t-1}^{j}) + \hat{z}_{t-1|t-1}^{j} +  q^{j-1}(x_{t+N-1|t-1}^{j}).\]
Also, according to equation (\ref{eq:comparision_overall_cost}), \[\mathcal{J}_{t\to t+N}^{j}(x_t^{*,j}) \leqslant \mathcal{J}_{t\to t+N|t-1}^{j}(x_{t}^{*,j}).\]
From the last two inequalities we conclude that
\begin{multline} \label{eq:theorem_asym_stab_final}
    \mathcal{J}_{t\to t+N}^{j}(x_t^{*,j}) - \mathcal{J}_{t-1\to t+N-1}^{j}(x_{t-1}^{*,j}) \leqslant \\ - \ell(x_{t-1|t-1}^{j} , u_{t-1|t-1}^{j}) - \hat{z}_{t-1|t-1}^{j} - q^{j-1}(x_{t+N-1|t-1}^{j}) \\< 0, \quad \forall t \geqslant 1, \quad \text{and} \quad \forall j \geqslant 1.
\end{multline}
This completes the proof of asymptotic stability of the equilibrium point $x_F$.  \hfill  $\filledmedsquare$


\begin{thm}\label{theorem:optimality_different_iter}
In the DMPC scheme with given system (\ref{eq:nldynamics}), cost function (\ref{eq:main_DMPC_concise_model}), and constraints (\ref{eq:DMPC_dynamic1}) - (\ref{eq:DMPC_xu_consts0}), and a feasible trajectory 
\normalfont $\textbf{x}^{*,j-1}$
at iteration $j-1$, 
\begin{equation}
    \mathcal{J}_{0\to\infty}^{*,j}(x_S) \leqslant \mathcal{J}_{0\to\infty}^{*,j-1}(x_S), \quad \forall j \geqslant 1
\end{equation}
the next trajectory $\textbf{x}^{*,j}$ generated by DMPC has an overall trajectory cost, $\mathcal{J}_{t\to\infty}^{*,j}(x_S)$, not worse than $\mathcal{J}_{t\to\infty}^{*,j-1}(x_S)$
\end{thm}

\textit{Proof}:
Assume that, at iteration $j$, the trajectory $\textbf{x}^{*,j-1}$ is available for an overall cost of $\mathcal{J}_{0\to\infty}^{*,j-1}(x_S)$. It is desirable to show that, according to model (\ref{eq:main_DMPC_concise_model}) and equation (\ref{eq:comparision_overall_cost}), DMPC will generate trajectory $\textbf{x}_{0:N}^j$ (trajectory blue) which is not worse than $\textbf{x}^{*,j-1}$, $\mathcal{J}_{0\to\infty}^{j}(x_S) \leqslant \mathcal{J}_{0\to\infty}^{*,j-1}(x_S).$ Positive definiteness of $z$ and $h$ indicates that at different time steps, $t$, in iteration $j$
\begin{equation}
    \mathcal{J}_{t\to t+N}^{j}(x_t^{*,j}) \leqslant \mathcal{J}_{t-1\to t+N-1}^{j}(x_{t-1}^{*,j}), \quad \forall t \geqslant 1.
\end{equation}
Also, according to equation (\ref{eq:theorem_asym_stab_final}), for $t=1$
\[\mathcal{J}_{0\to N}^{j}(x_{S}) \geqslant \mathcal{J}_{1\to N+1}^{j}(x_1^{*,j})+ \ell(x_S , u_{0}^{*,j}) + \hat{z}_{0}^{*,j} + q^{j-1}(x_{N|0}^{j})\]
for $t=2$, \[\mathcal{J}_{1\to N+1}^{j}(x_1^{*,j}) \geqslant\] \[\mathcal{J}_{2\to N+2}^{j}(x_2^{*,j})+ \ell(x_1^{*,j} , u_{1}^{*,j}) + \hat{z}_{1}^{*,j} + q^{j-1}(x_{N+1|1}^{j})\]
until $t \rightarrow{\infty}$, in which the system converges to $x_F$. Summing up these inequalities results in 
\[\mathcal{J}_{0\to N}^{j}(x_S) \geqslant \sum_{k=0}^{\infty} \left[ \ell(x_k^{*,j} , u_{k}^{*,j}) + \hat{z}_{k}^{*,j} + q^{j-1}(x_{k+N|k}^{j}) \right].\]
The right-hand side of this inequality is the sum of all stage costs of optimal trajectory generated at iteration $j$ \\
$\mathcal{J}_{0\to \infty}^{*,j}(x_S) = \sum_{k=0}^{\infty} \left[ \ell(x_k^{*,j} , u_{k}^{*,j}) + \hat{z}_{k}^{*,j} + q^{j-1}(x_{k+N|k}^{j}) \right],$
which yields the following inequality
\begin{equation} \label{eq:theorem_optimality_2}
     \mathcal{J}_{0\to N}^{j}(x_S) \geqslant \mathcal{J}_{0\to \infty}^{*,j}(x_S).
\end{equation}
From the last two inequalities we can easily conclude that
\begin{equation} \label{eq:theorem_optimality_final0}
     \mathcal{J}_{0\to\infty}^{*,j-1}(x_S) \geqslant \mathcal{J}_{0\to N}^{j}(x_S) \geqslant \mathcal{J}_{0\to \infty}^{*,j}(x_S),
\end{equation}
which shows, the overall cost of trajectories does not increase by the number of iterations
\begin{equation} \label{eq:theorem_optimality_final1}
     \mathcal{J}_{0\to \infty}^{*,j}(x_S) \leqslant \mathcal{J}_{0\to\infty}^{*,j-1}(x_S), \quad \forall j \geqslant 1,
\end{equation}
and the proof is complete.   \hfill  $\filledmedsquare$

%% file: implementation.tex
According to equation (\ref{eq:candidate_termina_states_set_0}), the DMPC algorithm needs the controllable set $\mathcal{S}_t^j$ at time step $t$ and iteration $j$ to select the best predicted terminal state. However, calculating such a set is a time consuming process and because it has to be executed for every state in different time steps $t$ of iteration $j$, it would affect the overall running time significantly. In this section we propose a technique to avoid this volume of unnecessary calculations. We assume that assumption (1) holds and there is a feasible trajectory from initial state $x_S$ to the equilibrium point $x_F$ which is given as $\textbf{x}^0$ and $\textbf{u}^0$. The main idea of this approach is that the algorithm will be given all of the states of the trajectory generated at the previous iteration, $\textbf{x}^{*,j-1}$, as terminal candidate states
\[
\mathcal{S}_t^j = \bigcup_{t=0}^{\infty}{x}_t^{*,j-1}.
\]
The algorithm selects the best predicted terminal state in this set from its current state ${x}_t^{*,j}$ using the following integer programming optimization model: 
\begin{subequations}\label{eq:Implem_DMPC_generic-model}
    \begin{align}
        J_{t\to t+N}^{j}(x_t^{*,j}) &=\min_{\textbf{u}_{t:t+N}^{j}} \sum_{k=t}^{t+N-1} \ell(x_{k|t}^{j} , u_{k|t}^{j}) \notag \\ &  + (N+1)\sum_{r=1}^{|Q_t^j|} \xi_r q^{j-1}(x_r^{*,j-1}) \label{eq:Implem_DMPC_generic-objective}\\
        \text{s.t.}\ & x_{k+1|t}^{j} = f(x_{k|t}^{j},u_{k|t}^{j})\quad \forall k \label{eq:Implem_DMPC_dynamic1}\\
        & x_{t|t}^{j} = x_t^{*,j} \label{eq:Implem_DMPC_init} \\
        & x_{t+N|t}^{j} = \sum_{r=1}^{|Q_t^j|} \xi_r x_r^{*,j-1} \label{eq:Implem_DMPC_terminal_state}\\
        & \sum_{r=1}^{|Q_t^j |} \xi_r = 1\label{eq:Implem_DMPC_just_one_state}\\
        & \xi_r \in \{0,1\}, \ \forall r = \{1,\dots, |Q_t^j|\} \label{eq:Implem_DMPC_binary_xi}\\
        & x_{k|t}^{j} \in \mathcal{X}, \ u_{k|t}^{j} \in \mathcal{U},\quad \forall k, \label{eq:Implem_DMPC_xu_consts0}
    \end{align}
\end{subequations}
where $\ell(x_{k|t}^{j}, u_{k|t}^{j})$ is stage cost and, in the second term of the cost function, $q^{j-1} \in Q_t^j$. $Q_t^j$ is the cost-to-go vector of terminal states in the set $\mathcal{S}_t^j$, which will be updated based on the current state ${x}_t^{*,j}$. However, in the beginning, the algorithm starts with $Q_t^j= \bigcup_{r=0}^{\infty}q^{j-1}(x_r^{*,j-1})$. We define a binary decision variable $\xi_r$ associated with each terminal state in the previous trajectory $\textbf{x}^{*,j-1}$. $\xi_r$ takes value one if the controller selects $r^{th}$ state from $\textbf{x}^{*,j-1}$ as the desirable terminal state and assigns value zero to other states; see constraint (\ref{eq:Implem_DMPC_terminal_state}). Also, using constraint (\ref{eq:Implem_DMPC_just_one_state}) we enforce the model to select only one state. The output of this model is given by $\overbar{\textbf{x}}_{t:t+N|t}^j$ and $\overbar{\textbf{u}}_{t:t+N|t}^j$.

Assume that the best terminal state selected by this model is $x_{\iota}^{*,j-1}$. Because the model has not considered $\sum_{k=t}^{t+N-1}\hat{z}_{k|t}^{j}$, the algorithm calls the available exogenous black-box system to calculate this value for the obtained trajectory $\overbar{\textbf{x}}_{t:t+N|t}^j$. Therefore, using equation (\ref{eq:seperated_model_data_cost}), $\overbar{\mathcal{J}}_{t\to t+N}^{j}(x_t^{*,j})$ can be found easily. After finding the overall trajectory cost from current state $x_t^{*,j}$ that passes through terminal state $x_{\iota}^{*,j-1}$, the algorithm updates the cost-to-go of state $x_{\iota}^{*,j-1}$ in set $Q_t^j$ from $q^{j-1}(x_{\iota}^{*,j-1})$ to $q^{j-1}(x_{\iota}^{*,j-1}) + \sum_{k=t}^{t+N-1}\hat{z}_{k|t}^{j}$. The algorithm keeps recording the index number of updated terminal states of set $Q_t^j$ in $I$. 

To calculate the complexity of the algorithm at iteration $j$, assume that at each iteration of Branch and Bound relaxation, the algorithm solves a convex quadratic model. Using the Interior Point Method (IPM), the computational complexity to find $\epsilon-$scale optimum for a quadratic model is polynomial in the size of the optimization model ($n'$) and required accuracy ($\epsilon$), i.e. $O(n'log1/\epsilon)$ \cite{ye1989extension}. The relaxation is implemented over the binary decision variables $\xi_r, \ \forall r$ defined for each terminal state in set $Q_t^j$. If the number of these candidate states is $T$, the worst-case number of iterations of the B\&B algorithm is exponential $O(2^T)$. On the other hand, the size of the model with time horizon $N$ is $(n+m)N$ at each time step $t$. In the worst case, all of the candidate states are tried to find the optimal candidate terminal state, which results in computational complexity of $O(2^T(n+m)NTlog1/\epsilon)$. The exponential part is dominant and yields in $O(2^T)$.

%% file: example.tex
\begin{figure}
    \centering
    \hspace{-10pt}\includegraphics[width=0.85\linewidth]{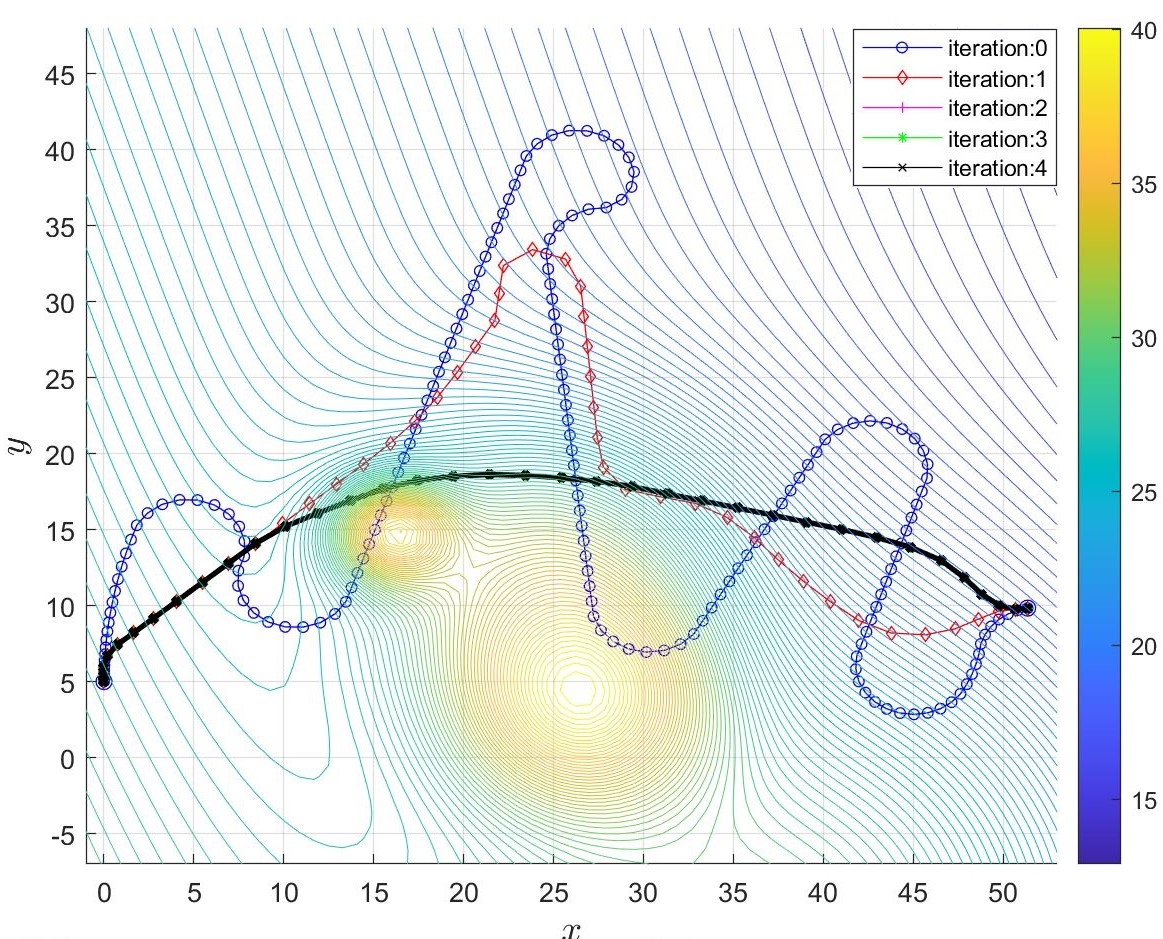}
    \caption{The contour plot of unknown non-convex cost function, and local optimal trajectory generated by DMPC. The contours are totally unknown to the controller. 
    }
    \label{fig:p1_DMPC_traj}
\end{figure}

We apply the proposed DMPC algorithm on the motion planning of an autonomous vehicle with a kinematic bicycle model in an inertial frame \cite{kong2015kinematic}. $\hat{z}$ is an unknown function and it is assumed that, given a trajectory, there is a black-box system that can predict its outputs and pass these to the controller. An example application of such a setting (see Fig.~\ref{fig:p1_DMPC_traj}) involves motion planning in an environment with regions that have different cost values, where the associated cost of selected states can be predicted by a machine learning-based black box. In motion planning, such black-box variables could include predictions of other agents' states or simply a region with uneven terrain or a potentially dangerous zone for a robot. The infinite time optimal control problem is defined according to  model (\ref{eq:main_model_generic_entire_model}), where $f(\mathrm{x}_t, \mathrm{u}_t)$ is defined as follows:
\begin{subequations} \label{eq:nonlinear_dynamics_vehicle}
    \begin{align}
        &\dot{x}_t = v_t \ cos(\psi_t + \beta_t)\\
        &\dot{y}_t = v_t \ sin(\psi_t + \beta_t)\\
        &\dot{\psi}_t = \frac{v_t}{l_r} \ sin(\beta_t)\\
        &\dot{v}_t = a_t.
    \end{align}
\end{subequations}
The state and control input vectors are $\mathrm{x}_t = [ {x}_{t} \ {y}_{t} \ {\psi_{t}} \ {v}_t]^T $,  $\mathrm{u}_t = [\delta_t \ a_t]^T $, respectively. $x_t$ and $y_t$ are the coordinates of the center of mass of the vehicle, $\psi_t$ is the heading angle, and $v_t$ is the velocity of the vehicle at time step $t$. $l_f$ and $l_r$ show the distance of the center of the mass from the front and rear axles, respectively. $\beta_t  = tan^{-1}(\frac{l_r}{l_f + l_r} \ tan(\delta_t))$ is the angle between the current velocity vector of the center of mass and the longitudinal axis of the vehicle. The control input vector $\mathrm{u}_t$ is composed of the steering angle $\delta_t$ and the acceleration $a_t$ that is defined for the center of mass in the same direction as $v_t$.

\begin{figure}[bp]
    \centering
    \hspace{-10pt}\includegraphics[width=0.85\linewidth]{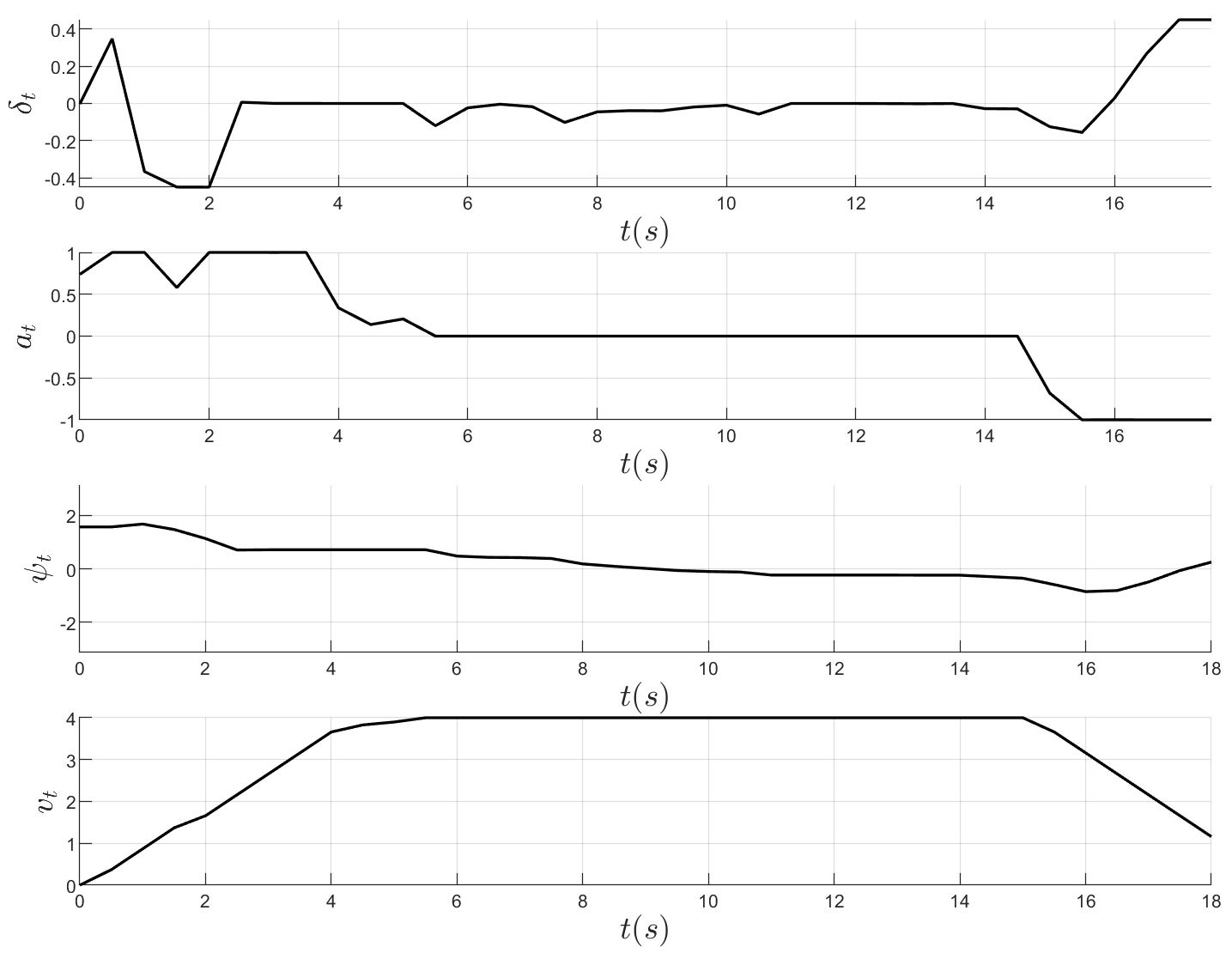}
    \caption{Control inputs and states in the steady state.}
    \label{fig:p1_DMPC_stat}
\end{figure}

The upper and lower bounds of the state and control vectors are $\mathrm{x}_{min} = [-\infty \ -\infty \ 0 \ 0]^T$, $\mathrm{x}_{max} = [+\infty \ +\infty \ 2\pi \ 4]^T$, $\mathrm{u}_{min} = [-\frac{\pi}{7} \ -1]^T$ and $\mathrm{u}_{max} = [\frac{\pi}{7} \ 1]^T$. The equality constraint representing initial state $\mathrm{x}_{0}$ is assumed to be $\mathrm{x}_{S} = [ 0 \ 5 \  \frac{\pi}{2} \ 0]$. Function $J_{0 \rightarrow{\infty}}(\mathrm{x}_0)$ shows the overall cost imposed to the controller to steer the system from initial state $\mathrm{x}_0$ to final state $\mathrm{x}_F = [51 \ 10 \ \frac{\pi}{10} \ 1.1]^T$. The stage cost $h(.,.)$ is defined as a quadratic function $h(\mathrm{x}_{t}, \mathrm{u}_{t}) = (\mathrm{x}_{t} - \mathrm{x}_{F} )^TP(\mathrm{x}_{t} - \mathrm{x}_{F} ) + \mathrm{u}_{t}^T{R}\mathrm{u}_{t}.$ The tuning matrices of the cost function are $P = diag[1 \ 1 \ 0.1 \ 0.1]$ and $R=diag[0.01 \ 0.01]$. 
In this example, the DMPC controller is expected to improve the given initial trajectory (blue circle trajectory in Figure~{(\ref{fig:p1_DMPC_traj})}) in the presence of an unknown cost function. The controller will use the most recently generated trajectory to converge to an optimal trajectory.
The algorithm will stop if $\sum_{t=0}^{\infty} |\mathrm{x}_{t}^{j} - \mathrm{x}_{t}^{j-1}| < 10^{-4}$. Also, the time step and time horizon is assumed to be $0.5$ second and $N=12$, respectively for this problem. We used ACADO Code Generation tool~{\cite{Houska2011b}} with MATLAB to solve this problem, and DMPC converged after 4 iterations (trajectories 2 and 3 are very close to the optimal solution that makes them invisible in the figure).Figure~{(\ref{fig:p1_DMPC_traj})} and ~{(\ref{fig:p1_DMPC_stat})} depict the generated trajectories $\textbf{x}^{*,j} \ \forall j \geqslant 0$, and optimal steering angle and acceleration/deceleration as control inputs, velocity and heading angle at different time steps.


Reinforcement learning (RL) is a natural candidate for comparison, but these approaches typically require a large number of interactions with the unknown system/function to learn controllers, which is a practical limitation in real cases, such as robots, where these number of interactions can be impractical, unsafe, and time-consuming~{\cite{deisenroth2013gaussian}}. In this group of applications Gaussian Process-based MPC outperforms the RL approaches, so we compare the performance of the DMPC with state-of-the-art GP methods ~{\cite{deisenroth2011pilco,kamthe2018data}}. We consider a Gaussian Process setting where we seek deterministic control inputs $\textbf{u}_{t}$ that minimize the cost function of the following finite time optimal control problems, which will be solved in a receding horizon fashion until reaching the terminal state \[\min_{\textbf{u}_t} \Big\{ J_{t \rightarrow t+N}(x_t) + \sum_{k=t}^{t+N} \mathbb{E}_{x_{k|t}}[\hat{z}(x_{k|t})] \Big\},\]
where $J_{t \rightarrow t+N}(x_t)$ denotes the conventional stage cost and $\mathbb{E}_{x_{k|t}}[\hat{z}(x_{k|t})]$ denotes the expected data-driven cost at time step $k$ calculated at time $t$. To implement the GP we define the training input and target data to be $\tilde{\textbf{x}} = [x \ y]^T$ and $\tilde{z}$ respectively.  We refer the reader to~{\cite{deisenroth2011pilco,kamthe2017data}} for details of the PILCO algorithm.
We use the same values of the parameters such as time horizon, time step, etc. However, without a decent reference trajectory this approach (PILCO) that is adopted from~{\cite{deisenroth2011pilco}} cannot find the optimal trajectory that drives the system to the terminal state. The reason for this result is that the MPC uses a naive approach (quadratic Euclidian distance from the equilibrium point) at each iteration to estimate the cost of the terminal state. Therefore, a reference trajectory is necessary for this approach, but it may be hard to compute such a trajectory. Alternatively, DMPC does not need any reference trajectory, and like RL, calculates a cost-to-go value for available states in the terminal set but in fewer trials than RL.


After adding a reference trajectory \cite{jafarzadeh2018exact} to the cost function and training the model with 5600 training samples, PILCO could solve the problem, whereas DMPC needs less than 2900 data samples, half the running time, and no reference trajectory. Another downside of using GP is that, even if the system has linear dynamics, adding such an estimation of $\hat{z}$ to the cost function will make the model non-convex. Such a result is not desirable in terms of running time and solution quality. Applying DMPC in this context results in a MILP model, which can be solved efficiently using off-the-shelf solvers such as CPLEX, Gurobi, etc.

%% file: conclusion.tex
In this work, a Data-and Model-driven Predictive Control (DMPC) algorithm is presented to solve a model predictive control problem in which there is a function in the performance index or constraints that (a) is unknown to the controller and (b) is interdependent with the decision variables (state and control vector) of the MPC. The controller is designed to exploit an existing, exogenous data-driven system such as a black-box deep learning model, along with model predictive control to find the optimal sequence of control inputs. To solve this problem, a controller is developed that conceptually borrows from iterative learning controller but is intended for non-iterative or nonrepetitive tasks.
The algorithm starts from an initial arbitrary trajectory and it is proven that the algorithm will find a feasible trajectory in each subsequent iteration, and the trajectory at each iteration is guaranteed to be no worse than the previous iteration. DMPC is effective with very little data and converges in only a few iterations. We provided an infinite time horizon optimal control example, in which the controller should drive a nonlinear system from an initial state to an equilibrium point where the environment is an uneven surface with an unknown non-convex shape. 